\documentclass[11pt,a4paper,oneside]{article}
\usepackage[utf8]{inputenc}
\usepackage[T2A,T1]{fontenc}
\usepackage{titlesec}
\usepackage{amsmath}
\usepackage{amsfonts}
\usepackage{mathtools}
\usepackage{amssymb}
\usepackage{graphicx}
\usepackage{subcaption}
\usepackage{amsthm}
\usepackage{setspace}
\usepackage{pgfplots}
\usepackage{fullpage}
\usepackage{exptex}
\usepackage{expthmi}
\usepackage{cite}
\usepackage{chicago}
\usepackage{rotating}
\usepackage{changepage}
\rmfamily
\makeindex
\makeatletter
\titleformat{\section}
{\normalfont\large\bfseries}{\thesection}{1em}{}

\begin{document}
\title{Intellectual and social similarity among scholarly journals: an exploratory comparison of the networks of editors, authors and co-citations}
\author{Alberto Baccini\footnote{Department of Economics and Statistics, University of Siena}, Lucio Barabesi\footnote{Department of Economics and Statistics, University of Siena}, Yves Gingras\footnote{Université du Quebec, Montreal}, Mahdi Khelfaoui\footnote{Université du Quebec, Montreal}}
\author{Alberto Baccini$^1$ \footnote{Alberto Baccini is the recipient of a grant by the Institute For New Economic Thinking Grant ID INO17-00015. The funders had no role in study design, data collection and analysis, decision to publish, or preparation of the manuscript. The authors are very grateful to the two reviewers for their very useful comments.}  \and Lucio Barabesi$^1$\and Mahdi Khelfaoui$^2$\and Yves Gingras$^2$}
\date{%
    $^1$Department of Economics and Statistics, University of Siena, Italy\\%
    $^2$Université du Quebec à Montreal, Canada\\[2ex]%
}
\maketitle

\section*{Abstract}
\newcommand{\pkg}[1]{{\normalfont\fontseries{b}\selectfont #1}}
\let\proglang=\textsf
\let\code=\texttt
This paper explores, by using suitable quantitative techniques, to what extent the intellectual proximity among scholarly journals is also a proximity in terms of social communities gathered around the journals. Three fields are considered: statistics, economics and information and library sciences. Co-citation networks (CC) represent the intellectual proximity among journals. The academic communities around the journals are represented by considering the networks of journals generated by authors writing in more than one journal (interlocking authorship: IA), and the networks generated by scholars sitting in the editorial board of more than one journal (interlocking editorship: IE). For comparing the whole structure of the networks, the dissimilarity matrices are considered. The CC, IE and IA networks appear to be correlated for the three fields.  The strongest correlations is between CC and IA for the three fields. Lower and similar correlations are obtained for CC and IE, and for IE and IA. The CC, IE and IA networks are then partitioned in communities. Information and library sciences is the field where communities are more easily detectable, while the most difficult field is economics.  The degrees of association among the detected communities show that they are not independent. For all the fields, the strongest association is between CC and IA networks; the minimum level of association is between IE and CC. Overall, these results indicate that the intellectual proximity is also a proximity among authors and among editors of the journals. Thus, the three maps of editorial power, intellectual proximity and authors communities tell similar stories.

\textbf{Keywords:} Interlocking editorship network; Gatekeeper; Interlocking authorship network; Co-citation network; generalized distance correlation; communities in network.
\newpage

\section{Introduction}
The main objects analyzed in this paper are scholarly journals and communities gathered around them. Scholarly journals have grown in relevance as outlet for communicating research results in the social sciences and humanities \cite{RN33}, following a trend that began in the natural sciences a century earlier \cite{RN24}. Over the last two decades, in the context of the publish-or-perish environment, where academic careers of scholars depend more and more on the "quality" of the journals in which they have published their articles, journals have gained a new importance as \textit{brands}\cite{RN6}. It is therefore hardly surprising that the interest of scientometric scholars for journals mainly focused on the building of indicators, such as the Impact Factor, to be used for evaluative purposes \cite{RN1}. The analysis of scholarly journals as social institutions of science appears a bit less developed.

Indeed, scholarly journals connect members of academic communities  \cite{RN9}. Editorial boards of journals constitute a first layer of such a community. They act as gatekeepers of science: they are, directly or indirectly, responsible for the refereeing processes, they decide which papers are worth publishing in their journals  \cite{RN32,RN25}. The stronger the link between the prestige of journals and the career advancement of scholars, the stronger the academic power exercised by the members of an editorial board. From this point of view, it is possible to consider editorial boars as engines  of academic power. 
A possible way for studying the role of editors consists in observing the presence of the same editors on the boards of different journals. The network of journals generated by the presence of the same person on the editorial board of more than one journal is called an Interlocking Editorship network (IE) \cite{RN2,RN3,RN4,RN5}. Thus, if two journals share the same persons on their editorial boards, it can be assumed that they have at least similar or complementary editorial policies, since they are managed by similar groups of scholars \cite{RN2}. From another perspective, editors have the power to drift the paper selection processes toward decisions favoring department colleagues, or disciples, and so on \cite{RN8,RN11}. In this sense the IE network can be used to try to identify some kind of favoritism in the refereeing processes \cite{RN7}, or for illustrating the self-referentiality of national communities of scholars \cite{RN3}. 

A second social community gathered around scholarly journals is constituted by the authors of the published  articles. While many studies exist about authorship and co-authorship, only a few are focused on the communities of authors of specific journals \cite{RN9}.
In turn, it is possible to work analogously to the IE network by considering the journal network generated by the scholars authoring papers in different journals. The network among journals generated by the crossed presence of the same authors in different journals could be called the Interlocking Authorship (IA) network. To the best of our knowledge, this kind of network has been rarely explored \cite{RN10,RN35,RN36}. In the IA network, the proximity between two journals can be considered proportional to the number of common authors. Such a  proximity is, in some sense, \textit{intellectual} since it is based on the choices made by authors on where to publish their papers, and on decisions of the editors to accept or not to publish those papers. The community of authors around a journal thus reflects to a certain degree the contents of the journal and the activity of the gatekeepers of the journal. If two journals are in proximity, it can be supposed that they have similar contents and that their  editorial policies are similar or complementary.

Scholarly journals contribute to the definition of the intellectual landscapes of research fields. Co-citation analysis is probably the best known instrument for studying the intellectual proximity among authors, papers and journals \cite{RN26}. For instance, if two authors are frequently cited together in many different papers, this suggest that these two persons are somehow intellectually connected by the topic or methodology of their work. Similarly, two different journals often cited together in the same paper suggest that these journals are connected. The more often they are cited together the stronger the link between these authors or journals. We thus obtain a network connecting the journals based on their being often cited together. Let us call this network CC as it is based on a different measure than those obtained through IE and IA.  

In this paper we consider the IE, IA, CC networks of journals summarily described above and we compare the degree of proximity of journals in the three networks. 
The first intuitive question is to what extent these three networks are similar. If two journals are well connected in the CC network, that is if they have a strong intellectual proximity, does a similar proximity exists in the IA or IE network? The basic idea is to explore to what extent the \textit{social} proximity among journals observed in the network of the editorial boards is similar to the social/intellectual proximity observed in the IA network and in the intellectual proximity in the CC network. 

This question is explored by considering the IE, IA and CC networks in three fields: economics (EC), statistics (STAT), and information and library science (ILS).  Two reasons justify the choice of the three fields. The first one is practical: for the three fields data on the editorial boards of journals were already available because they had been collected by two of the authors in a previous research project. Data on editorial boards have to be collected by hand. Hence, their availability is a big advantage. The second reason is that scholars in the three fields differ in the way they use scholarly journals as outlet for publishing research results. While in statistics journals articles are largely dominant, scholars in economics and in information and library sciences continue to write book chapters and books \cite{RN33}. Hence the similarity analysis considered three different scholarly communication contexts.

For each field we compare the three networks as a whole by using suitable statistical techniques. Subsequently, for each field, we partition the three networks in "communities of journals" and we analyze the coherence of these communities between pairs of networks. 

The paper is organized as follow. Section 2 contains the technical definitions used in the rest of the paper and the description of the methodology adopted for empirical analysis. Section 3 contains the outcome of the analysis. Section 4 discusses the results and Section 5 concludes by suggesting further steps of the present research.

\section{Journal networks data}

The journal networks considered here are all one-mode \cite{RN14}. In an IE network, nodes are scholarly journals and the edge between two journals indicates that at least one scholar sits in the board of the two. Each edge can be weighted by the number of common editors between the linked journals. Analogously, in the IA networks, the edges between journals are generated by common authors and the weight of the edge is the number of common authors. Finally in a CC network, the edge between journals is generated by the fact that the two journals are cited together at least in one article; the weight of the edge is the number of articles citing the two journals together.

We have constructed the three networks (IE, IA, CC) for the three fields for a total of nine networks. For IE networks, as anticipated, we used three existing databases,  each containing the journal editorial boards in a given year. Details on their collection and normalization can be found in the papers referenced below. Moreover, IA and CC networks were constructed by using Web of Science (WoS) data for a five years time-period starting from the year for which the IE was recorded.  The raw data for the nine one-mode networks can be downloaded from https://doi.org/10.5281/zenodo.3350797.

\begin{sidewaysfigure}
 \centering	
     \includegraphics [scale=0.5]{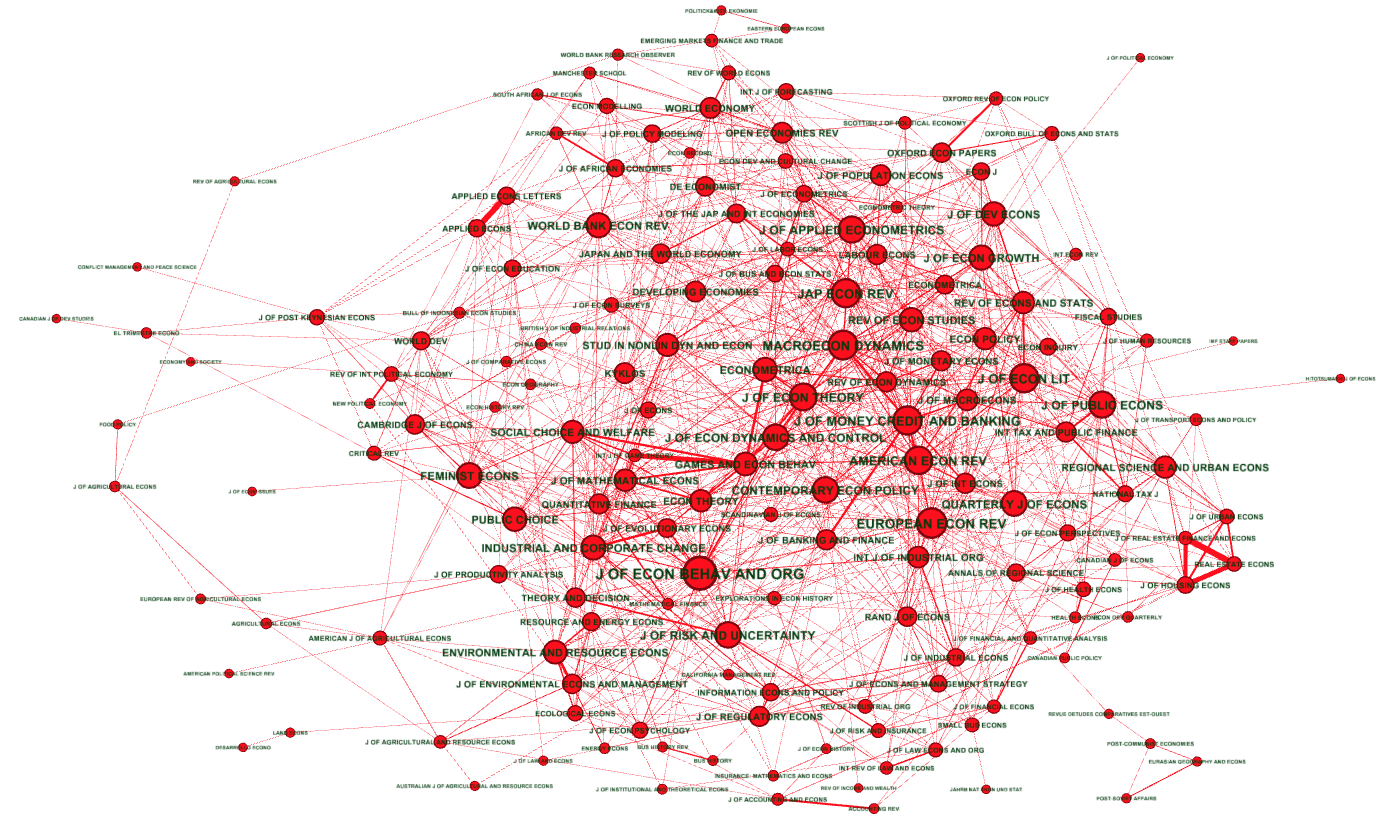} 
   \caption{Interlocking editorship network of economic journals}
   \label{fig:IE_ECON}
\end{sidewaysfigure}

\begin{sidewaysfigure}
 \centering	
     \includegraphics [scale=0.5]{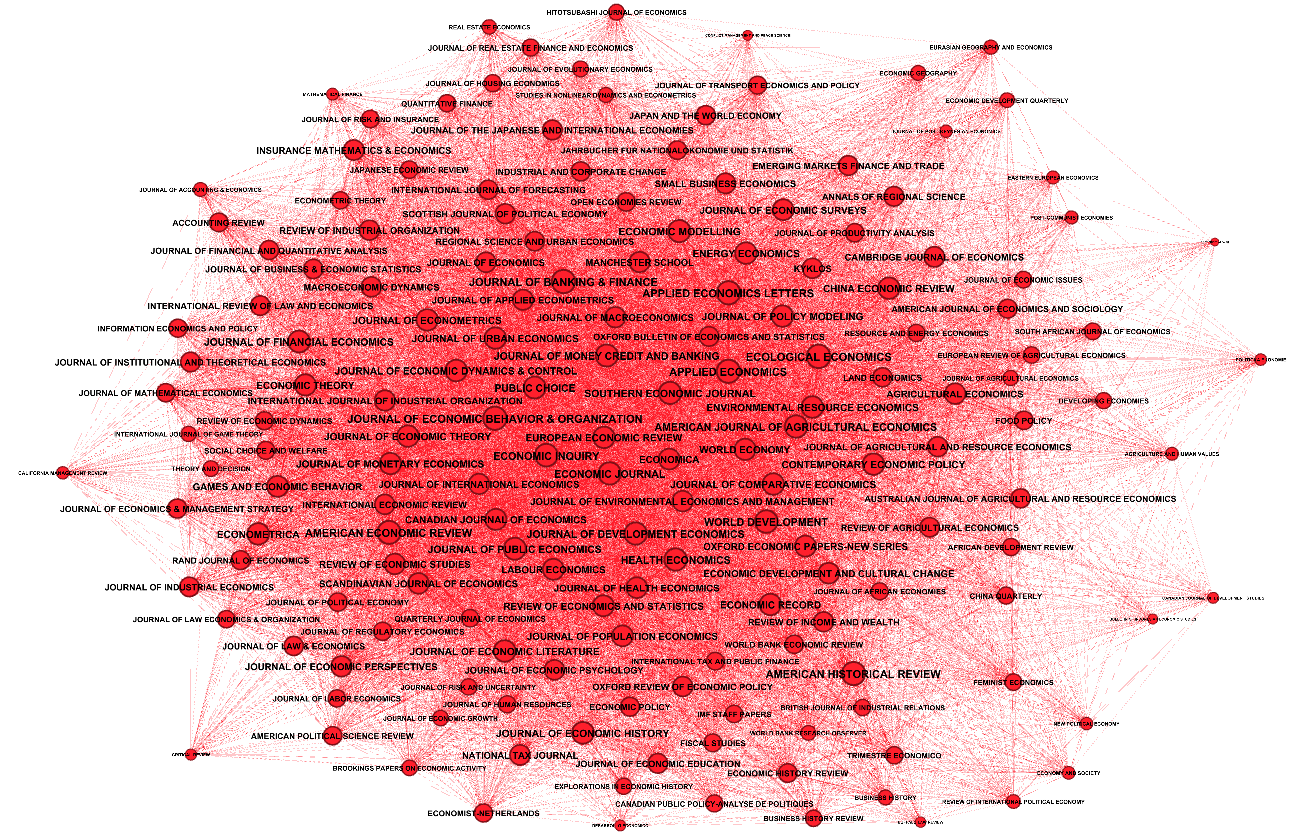} 
   \caption{Interlocking authorship network of economic journals}
 \label{fig:IA_ECON}
\end{sidewaysfigure}

\begin{sidewaysfigure}
 \centering	
     \includegraphics [scale=0.5]{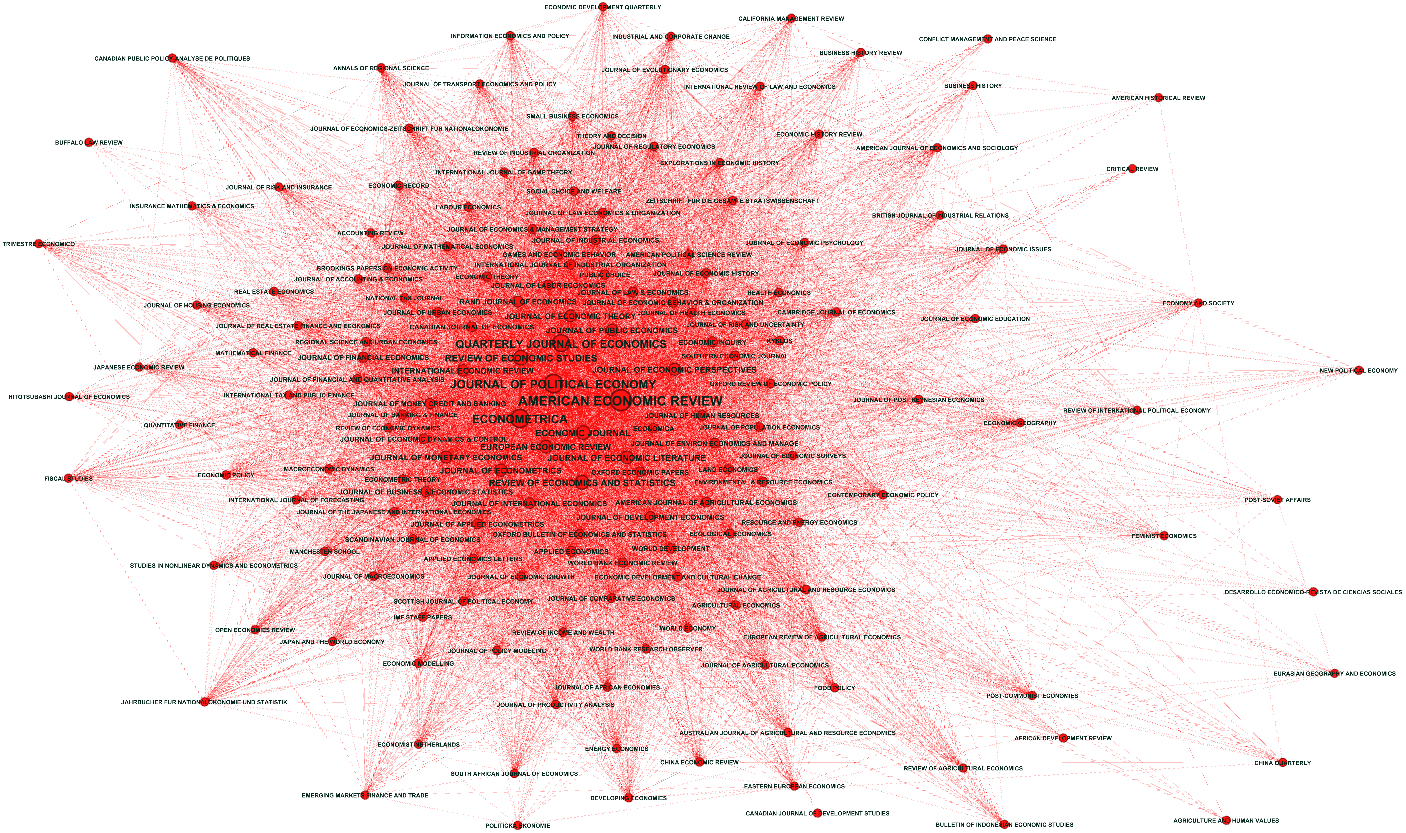} 
   \caption{Co-citation network of economic journals}
 \label{fig:CC_ECON}
\end{sidewaysfigure}

For economics, we considered a set of 169 journals listed in the EconLit database, and indexed in the \textit{Journal Citation Reports} for the year 2006. The IE network (Figure \ref{fig:IE_ECON}) was extracted from the database collected by Baccini and Barabesi \citeyear{RN21} for the year 2006. The IA (Figure \ref{fig:IA_ECON}) and CC (Figure \ref{fig:CC_ECON}) networks for economic journals were built on WoS data, by considering respectively the authors of and the references in the papers published in the journals in the years 2006-2010. 

\begin{sidewaysfigure}
 \centering	
     \includegraphics [scale=0.5]{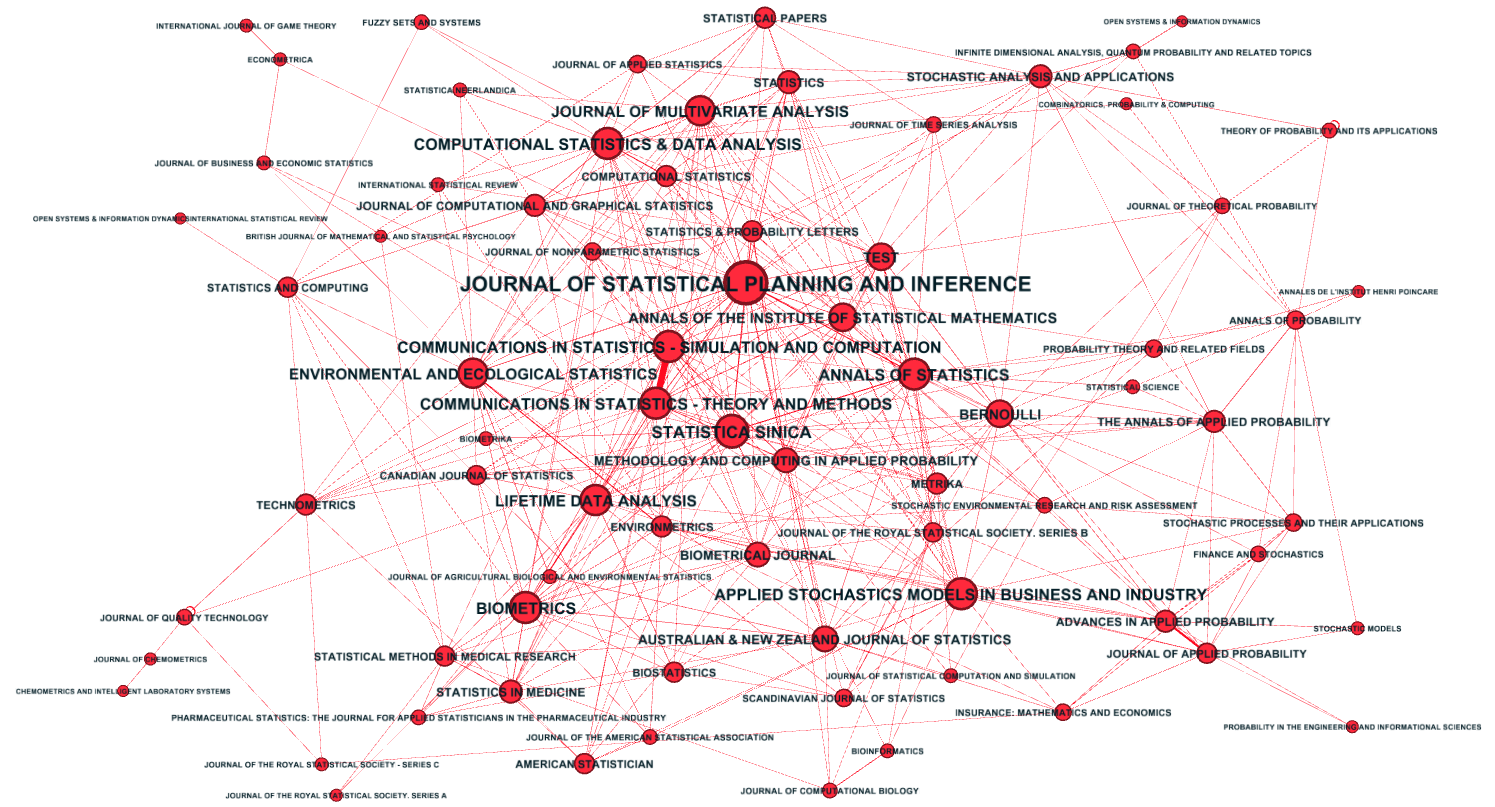} 
   \caption{Interlocking editorship network of statistical journals}
 \label{fig:IE_STAT}
\end{sidewaysfigure}

\begin{sidewaysfigure}
 \centering	
     \includegraphics [scale=0.5]{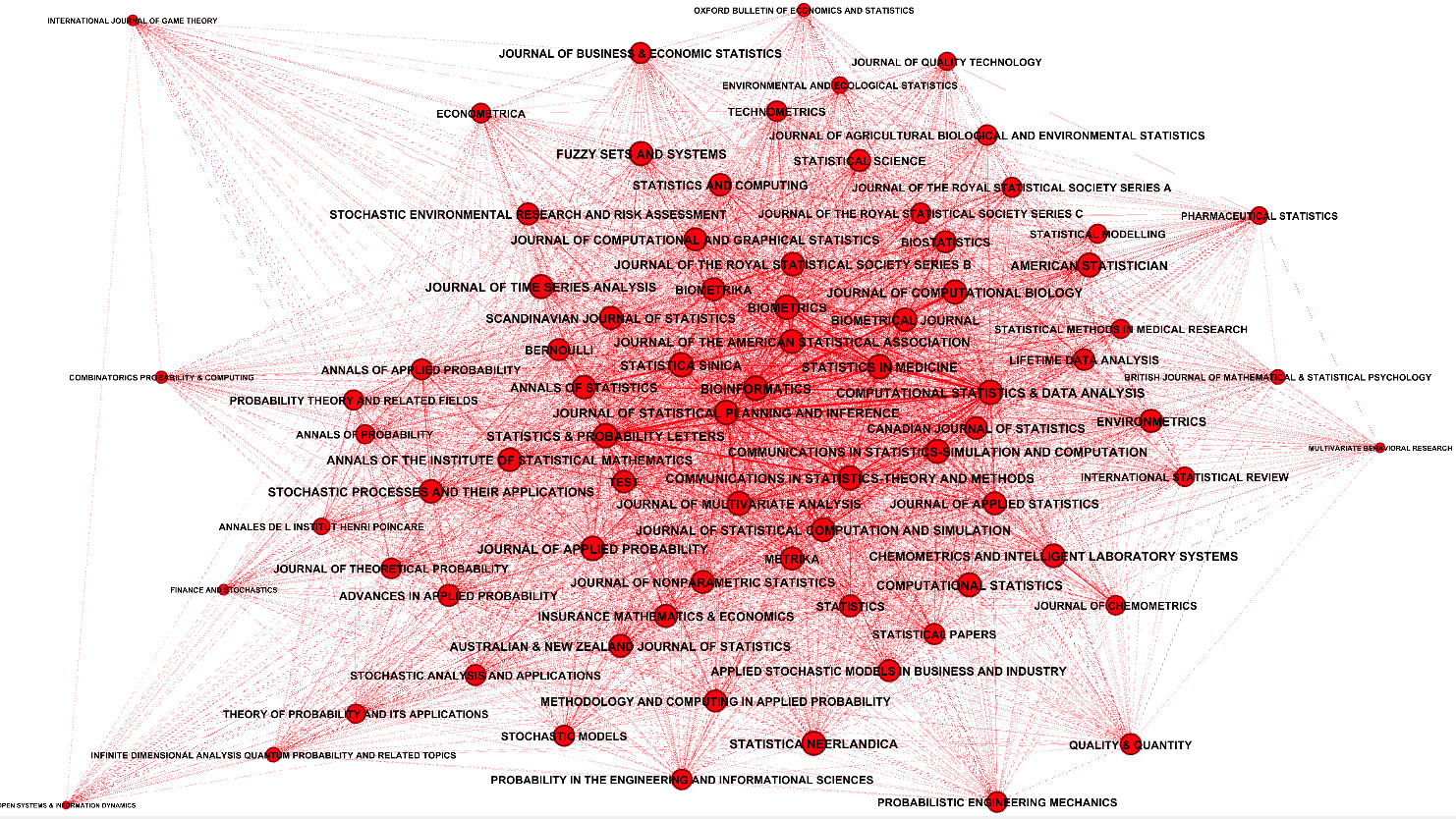} 
   \caption{Interlocking authorship network of statistical journals}
 \label{fig:IA_STAT}
\end{sidewaysfigure}

\begin{sidewaysfigure}
 \centering	
     \includegraphics [scale=0.5]{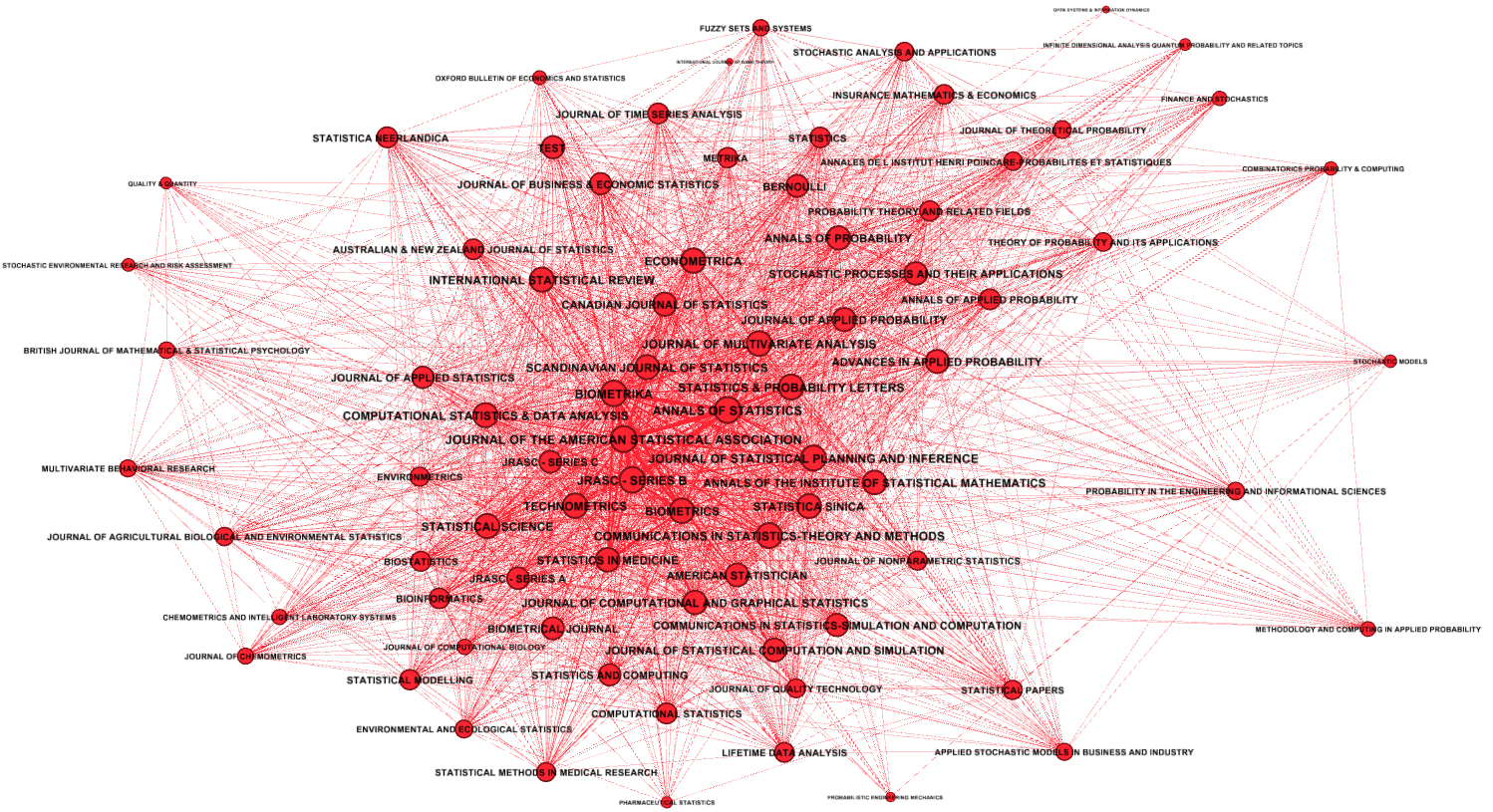} 
   \caption{Co-citation network of statistical journals}
 \label{fig:CC_STAT}
\end{sidewaysfigure}

For the field of statistics, the set includes the 79 journals listed in the category "Statistics and probability" of the \textit{Journal Citation Reports} for the year 2005. IE data (Figure \ref{fig:IE_STAT}) are the ones collected in Baccini, Barabesi and Marcheselli \citeyear{RN2} for the year 2006. Similarly, for the discipline of statistics, IA (Figure \ref{fig:IA_STAT}) and CC (Figure \ref{fig:CC_STAT}) networks were built using WoS data, by considering papers published in the years 2006-2010.

\begin{sidewaysfigure}
 \centering	
     \includegraphics [scale=0.5]{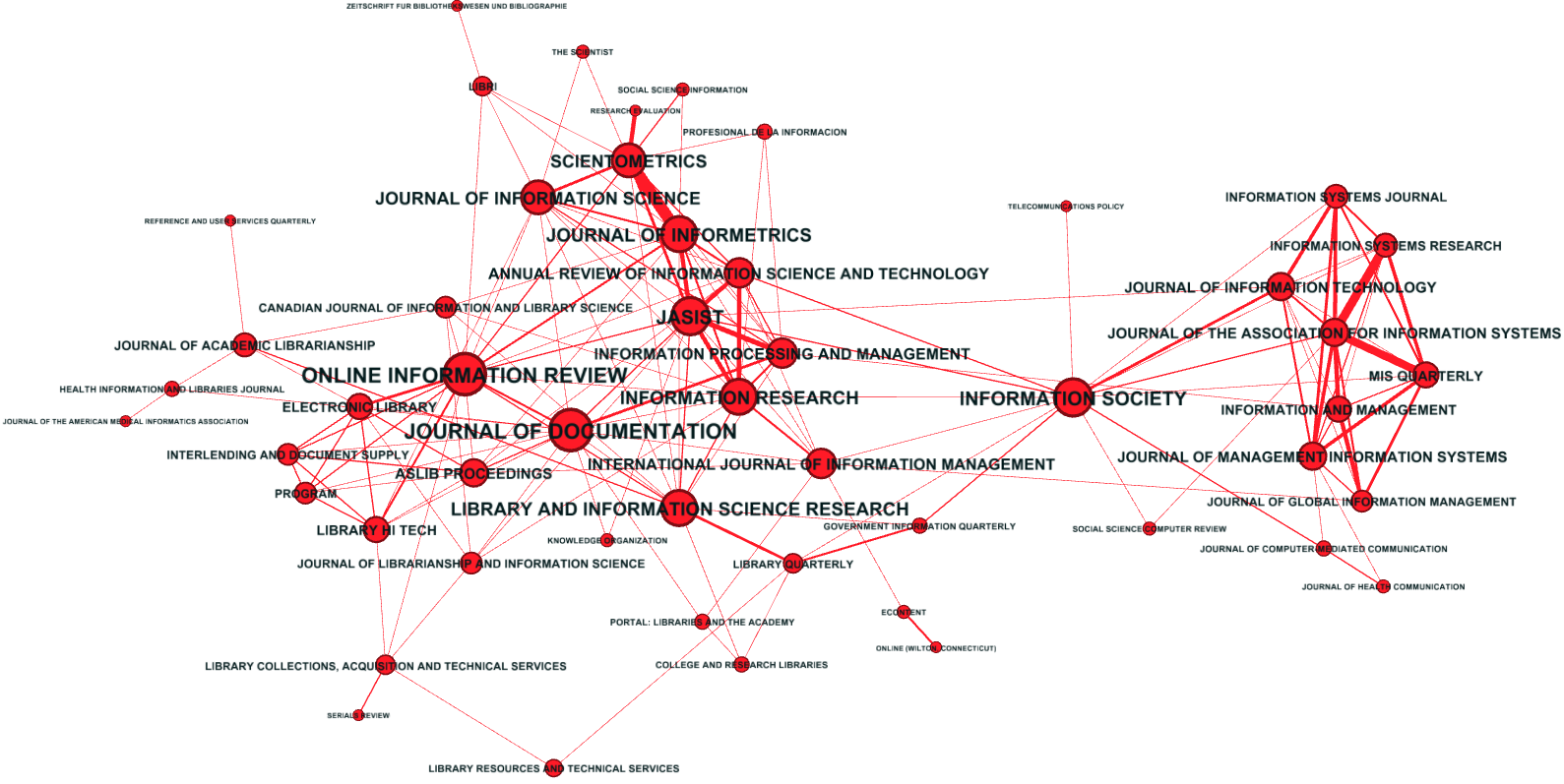} 
   \caption{Interlocking editorship network of information and library sciences journals}
 \label{fig:IE_ILS}
\end{sidewaysfigure}

\begin{sidewaysfigure}
 \centering	
     \includegraphics [scale=0.5]{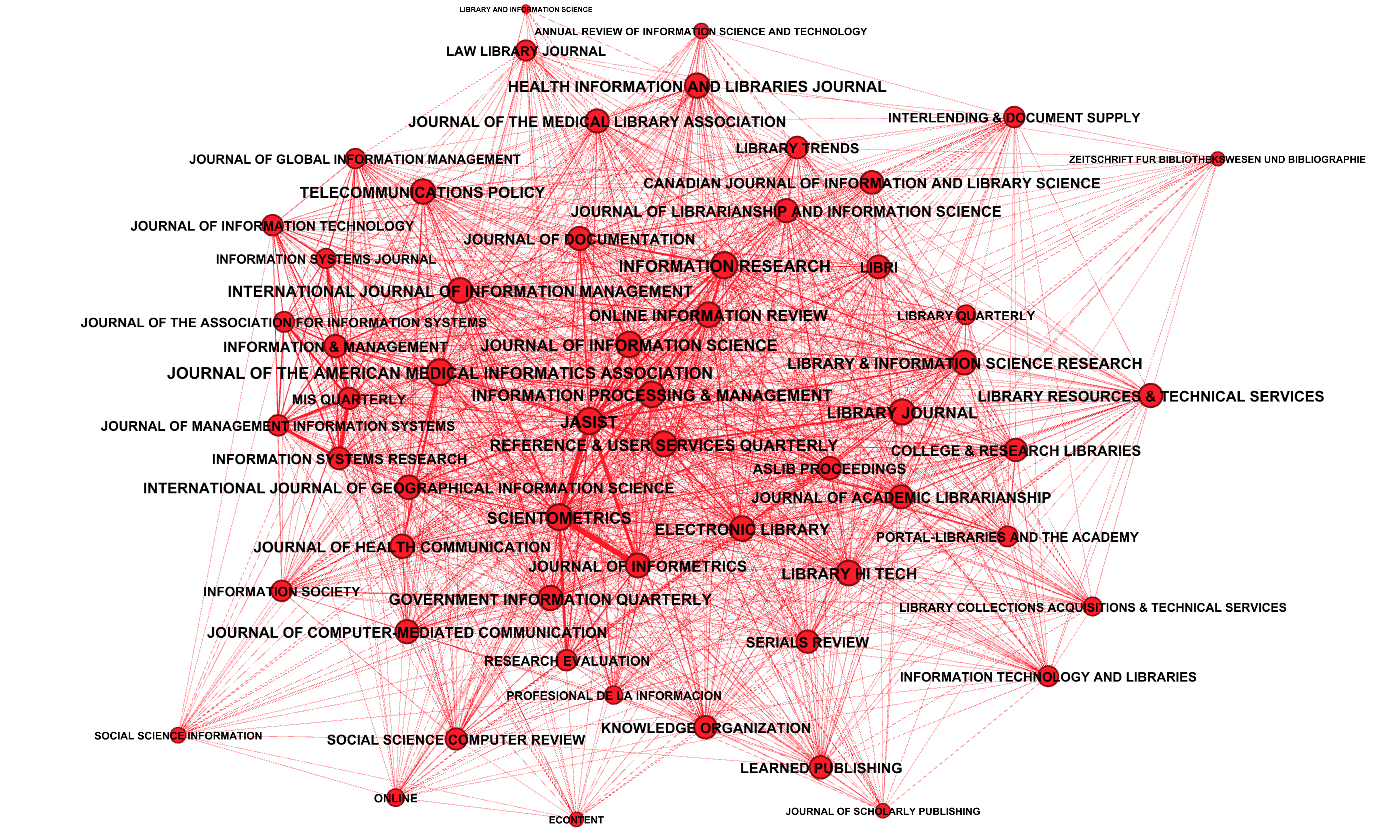} 
   \caption{Interlocking authorship network of information and library sciences journals}
 \label{fig:IA_ILS}
\end{sidewaysfigure}

\begin{sidewaysfigure}
 \centering	
     \includegraphics [scale=0.5]{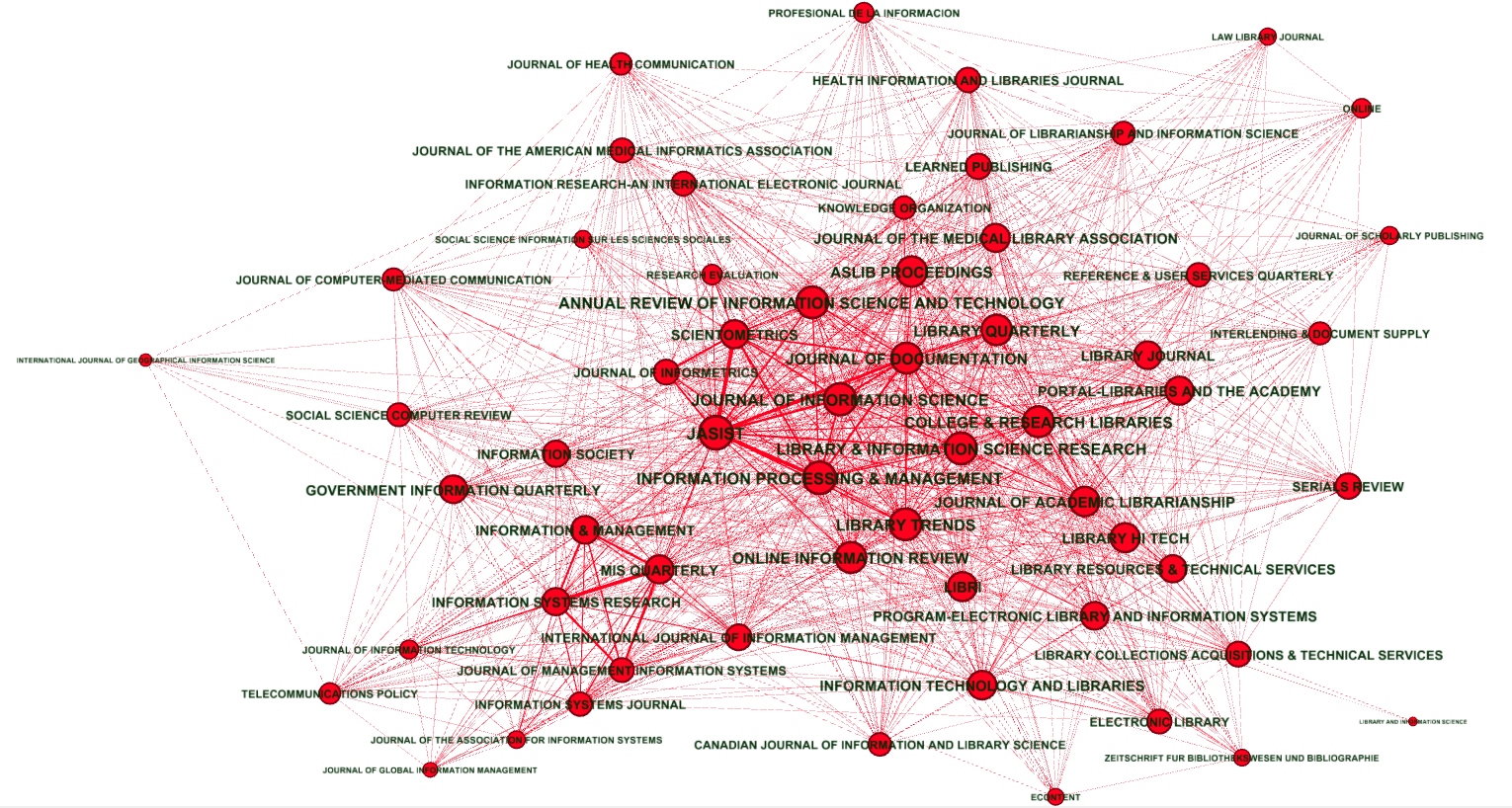} 
   \caption{Co-citation network of information and library sciences journals}
 \label{fig:CC_ILS}
\end{sidewaysfigure}

Finally, for the domain of information and library sciences, the set includes the 59 journals listed in the category "Information science and library science" of the \textit{Journal Citation Reports} for the year 2008. IE data (Figure \ref{fig:IE_ILS}) are the ones collected in Baccini and Barabesi \citeyear{RN4} for the year 2010. Again, IA (Figure \ref{fig:IA_ILS}) and CC (Figure \ref{fig:CC_ILS}) networks were built on WoS data, by considering papers published in the years 2010-2014. 

In Figures 1-9 the size of a node is proportional to its degree and the width of an edge is proportional to the value of the link. In the IE network, for example, the size of a node is proportional to the number of journals to which it is linked; the width of the link between two nodes is proportional to the number of their common editors. For each field, the visual comparison of the three networks is hardly informative. For instance, it is apparent that for all three fields, the IE networks are less connected than the IA and CC networks. Also, in the center of the networks there are not always the same journals; and a journal may have a different size in the three networks. We therefore need a better way of comparing networks.  

\section{Dissimilarities among networks}

For each network, it is possible to build a pseudo-measure of the distance among journals by calculating a matrix of dissimilarities. The Jaccard index was adopted as a dissimilarity measure (for more details on the Jaccard index, see e.g. Levandowsky and Winter \citeyear{RN22}). More precisely, if $A$ and $B$ represent the sets containing the members of the editorial boards of two journals, the Jaccard dissimilarity is defined as
\begin{equation}
J(A,B)=\frac{\mid A\cup B\mid - \mid A \ \cap B\mid}{\mid A\cup B\mid}\
\end{equation}

As an example, in the IE network, the similarity among journals is proportional to the number of common editors in their boards. Hence, the minimum dissimilarity $J(A,B)=0$ is reached when two journals have exactly the same editorial board, i.e. all the editors of a journal are also the editors of the other and vice versa. The maximum dissimilarity $J(A,B)=1$ is reached when two journals have no editors in common. 
In order to compare the three dissimilarity matrices arising from co-citation, editorial board and author networks for each discipline, we adopt the generalized distance correlation $R_d$ suggested by Omelka and Hudecov\a'a \citeyear{RN28} on the basis of the seminal proposal by Székely et al. \citeyear{RN30}. It should be remarked that such a correlation index avoids the drawbacks emphasized by Dutilleul et al. \citeyear{RN20} when the classical Mantel coefficient is assumed instead \cite{RN28}. Hence, we considered the three possible couples of networks and we computed the corresponding values of $\sqrt{R_d}$ for each discipline. It is worth noting that $R_d$ is somehow similar to the squared Pearson
correlation coefficient - and hence $\sqrt{R_d}$ should be interpreted as a
generalization of the usual correlation coefficient. More precisely, $R_d$ is
defined in the interval $[0,1]$, in such a way that values close to zero
indicate no or very weak association, while larger values suggest a stronger
association, which is perfect for $R_d=1$ - and similar considerations
obviously hold for $\sqrt{R_d}$ (for more details, see e.g. Omelka and Hudecov\a'a \citeyear{RN28}).
The generalized distance correlation was evaluated in the \proglang{R}-computing environment \cite{RN31} by using the \proglang{R} function \code{dcor} in the package \pkg{energy} \cite{RN29}.
These values of $\sqrt{R_d}$ are reported in Table 1. 

\begin{table}
\caption{Generalized distance correlations between networks}
\centering
\scriptsize
{}
{%
\begin{tabular}[t]{ccccc}
\hline
Networks&&Statistics&Information and library sciences&Economics\\
\hline
co-citation vs editor &$\sqrt{R_d}$&$0.5947$&$0.5386$&$0.5228$\\
\ &P-value&$0.00001$&$0.00058$&$0.00001$\\
&&&&\\
co-citation vs author&$\sqrt{R_d}$&$0.6431$&$0.6389$&$0.7518$\\
&P-value&$0.00001$&$0.00001$&$0.00001$\\
&&&&\\
editor vs author&$\sqrt{R_d}$&$0.5985$&$0.4969$&$0.5112$\\
&P-value&$0.00001$&$0.00382$&$0.00001$\\
\hline
\
\end{tabular}%
}{}
\end{table}

From the analysis of Table 1, the dependence between the considered dissimilarity matrices is apparent. Indeed, the observed values of $\sqrt{R_d}$ are greater than (or nearly equal to) the value $0.5$ for each combination of networks in the three disciplines. Moreover, the permutation test for assessing independence, as proposed by Omelka and Hudecov\a'a \citeyear{RN28}, was also carried out. The statistical details of the permutation test are rather involved, even if they are clearly explained by Omelka and Hudecov\a'a \citeyear{RN28}. Loosely speaking, the rationale behind the test stems from the fact that, under the null hypothesis of independence, the generalized distance correlation should not be affected by a random permutation of the rows and the corresponding columns of the “centred”
distance matrices. The permutation principle is widely adopted in order to carry out
nonparametric inference, since assumptions are minimal and practical implementation is
often straightforward (see e.g. Lehmann and Romano \citeyear{RN34}, Section 10). The permutation test of independence was in turn implemented by using the package \pkg{energy} \cite{RN29}. The significance of the test statistic was computed by means of the \proglang{R} function \textsf{dcov.test} (for more details Omelka and Hudecov\a'a \citeyear{RN28}). On the basis of the achieved P-values given in Table 1, the independence hypotheses can be rejected at the significance level $\alpha=0.01$.
Since the three statistical tests within each discipline are obviously dependent, we also consider the Bonferroni procedure in order to control the familywise error rate (for more details, see e.g. Bretz, Hotorn and Westfall \citeyear{RN19}). Thus, by assuming such a procedure and a global significance level given by $\alpha=0.01$, the marginal independence hypotheses may be rejected if the corresponding P-values are less than $\alpha/3=0.0033$, which is the case for all the considered tests - except the editorial board and author networks for information and library sciences.
However, it is worth remarking that - even in this case - the corresponding P-value is just slightly larger than the threshold. Hence, the co-citation, editorial board and author networks display structures which may be considered associated for each considered discipline - at least on the basis of the considered dissimilarity matrices.

\section{Correlations among communities of journals}

The proximity among journal networks can be explored by focusing on communities of journals. The first step consists in detecting communities inside each network; the second in verifying the degree of association between the communities detected in different networks of the same field. A non-overlapping community of nodes of a network is a set of nodes densely connected internally and only sparsely connected with external nodes. 
Each network is partitioned in communities by using the Louvain algorithm \cite{RN37} as implemented in the software  \proglang{Pajek} \cite{RN15}. It consists in the optimization of the modularity of the network \cite{RN39,RN38}. The quality of the partition is quantitatively measured by modularity values. Table 2 reports the values of modularities and the resolution parameters adopted for optimization. The resolution parameter is used to control the size of the communities detected; higher values of the parameter produce larger number of communities and viceversa. Table 2 also reports the number of communities detected.

\begin{table}
\caption{Main features of networks and communities}
\
\centering
\scriptsize

{}
\begin{tabular}{cccccccccc}
\hline 
 & \multicolumn{3}{c}{Statistics} & \multicolumn{3}{c}{Information and library sciences} & \multicolumn{3}{c}{Economics} \\ 
\
\ &IE&CC&IA&IE&CC&IA&IE&CC&IA\\
\hline
Density&0.121&0.671&0.91&0.0935&0.644&0.764&0.07&0.566&0.744 \\ 
\
Average degree&9.443&52.379&70.962&5.423&37.356&44.339&11.751&95.053&125.001 \\ 
\
Isolated journals&4&0&0&9&0&0&7&0&0 \\ 
\
Resolution&1&1&1&1&1&0.8&1&1&1 \\ 
\
Modularity&0.4&0.108&0.171&0.528&0.266&0.329&0.444&0.09&0.218 \\ 
\
n. communities&10&3&4&14&3&3&16&4&5 \\ 
\
n. non-isolated communities&6&3&4&5&3&3&9&4&5\\ 
\
E-I unweighted&-0.04&0.22&0.41&-0.425&0.201&0.2&0.108&0.322&0.435 \\ 
\
E-I weighted&-0.309&-0.141&-0.02&-0.651&-0.355&-0.328&-0.202&0.131&0.038 \\ 
\hline 
\end{tabular} 

\end{table}

For all the pairs of the networks inside each research field, the association between the resulting communities is then analyzed by using statistical techniques as available in \proglang{Pajek}  \cite{RN15}. All the indicators considered are adopted under an exploratory approach.
$\chi^2$ statistics provide an index aiming to assess the degree of independence of the partitions of each pair of networks. Cramér's $V$ is a measure of association giving a value between $0$ (no association)  and $+1$ (perfect association) \cite{RN17}. Rajski's coherence \cite{RN16} is presented in three variants, all defined in [0,1] range: a symmetrical version indicating the coherence between each pair of classification; and two asymmetrical versions called in Table 3 "Rajski's right" and "Rajski's left". When the communities in the IE-CC networks are considered, Rajski's left indicates the extent to which the first communities classification IE is able to predict the second communities classification CC; Rajski's right indicates instead the extent to which the second classification is able to predict the first. Finally, the adjusted Rand index measures the degree of association between partitions and is bounded between $\pm1$ \cite{RN18}. All indices are reported in Table 3.

For the three fields analyzed here, we observe that the IE is the least dense network and the network with the lowest average degree. For the three fields, the CC networks are in the intermediate position for density and average degree, and finally the IA networks have the highest values of density (0.91 for statistics) and average degree \cite{RN14}. 

In general, the community detection algorithm was more successful in sparser networks: for the three fields, the values of modularity are indeed the highest for the IE network, intermediate for CC and lowest for IA. In the IE networks many detected communities are actually isolated journals, i.e. journals with no common editors with other considered journals. In every case, the number of communities detected in the IE networks is always bigger than the number of communities detected in the other networks.

\begin{sidewaystable}
\caption{Association indexes between communities}
\
\centering
\scriptsize

{}
\begin{tabular}{cccccccccc}
\hline 
 & \multicolumn{3}{c}{Statistics} & \multicolumn{3}{c}{Information and library sciences} & \multicolumn{3}{c}{Economics} \\ 
&IE-CC&IE-IA&CC-IA&IE-CC&IE-IA&CC-IA&IE-CC&IE-IA&CC-IA\\
\hline
\ $\chi^2$ (d.f.)&68.59 (18)&118.88(27)&80.85(6)&75.45 (26)&80.51(26)&68.68(4)&199.58(45)&255.34(60)&207.29(12) \\ 
\
Cramér's V&0.659&0.708&0.715&0.799&0.826&0.763&0.627&0.615&0.639 \\ 
\
Rajski's&0.207&0.291&0.312&0.279&0.301&0.434&0.171&0.207&0.229 \\ 
\ 
Rajski's right&0.448&0.527&0.524&0.624&0.679&0.617&0.398&0.431&0.351 \\ 
\
Rajski's left&0.278&0.394&0.435&0.336&0.351&0.594&0.232&0.284&0.398 \\ 
\
Adj. Rand index&0.249&0.346&0.444&0.296&0.303&0.503&0.147&0.2&0.265\\ 
\hline 
\end{tabular} 

\end{sidewaystable}

Information and library sciences is the field where the communities are more easily detectable and more clearly defined, as shown by the highest modularity values and by the lowest values of the E-I indices (Table 2) \cite{RN15}. In particular for the IA network, communities were detected by adopting a resolution value of $0.8$. This resolution was preferred to the value of $1$ adopted for all the other networks, because the resulting communities exhibited better E-I indices\footnote{With a value of resolution of $1$ the IA network is partitioned in $4$ communities, with modularity $0.257$, E-I unweighted$=0.255$ and E-I weighted=$-0.083$.}. The E-I index was calculated as the difference between the number of edges within communities and the number of edges between communities; that difference is then divided for the total number of edges of the network. The weighted version of the index is calculated by considering the value of the edges.   The range of the index is between -1 (all edges are inside communities) and 1 (all edges are between communities). The $\chi^2$ values show that the detected communities for the three networks are not independent. The association between the partitions of communities as measured by Cramér's $V$ is high. The highest level of association as measured by the adjusted Rand's index is found between communities detected in the CC and IA networks.  Rajski's right indicates that the communities detected in the IE network predict well the communities detected in the other networks.  

The field of statistics is in an intermediate position: the values of modularity are very low for CC and IA networks, nevertheless the resulting partitions have negative values of the E-I weighted indices. Communities in the IE network are more easily detectable and more clearly defined than in the IA and especially in the CC network. Also in this field, the $\chi^2$ values show that the detected communities for the three networks are not independent. The association between the partitions of communities is a bit higher between CC and IA than for the other pairs of networks. Also in this case Rajski's right indicates that the communities of the IE network predict well the communities in the other two networks.

For the case of economics, community detection is particularly problematic, i.e. small changes of the value of the resolution parameter changed substantially the number of detected communities and the values of the indicators considered. For CC and IA, the community detection procedure results in very low values of modularity and in positive values of E-I. Only for the IE network the modularity is around 0.5 and the value of the E-I weighted index is less than zero. Also in this field the $\chi^2$ values show that detected communities for the three networks are not independent. The association between the partitions of communities is the lowest of the three fields analyzed in this paper. Rajski's right indicates also for economics that the communities in the IE network predict the communities in the other two networks, but the values of the index are the lowest of the three.

\section{Discussion and conclusions}

The main aim of this paper was to explore, by using suitable quantitative techniques,  to what extent the intellectual proximity among scholarly journals is also a proximity in terms of social communities gathered around the journals. 

For representing the intellectual proximity among journals we have used the CC network. For having information about the academic communities around journals, we have considered the networks of journals generated by authors writing in more than one journal as well as the networks generated by scholars sitting in the editorial board of more than one journal. 
The first step of the exploratory analysis consisted of comparing the whole structure of the networks on the basis of dissimilarity matrices. The CC, IE and IA networks appear to be associated for all the three considered fields. 
The second step consisted of partitioning the IE, IA and CC networks in communities and then in verifying the degree of association among the detected communities. The results of that analysis show that the communities detected in the three networks are not independent for the three research fields considered. The results of both approaches are coherent in showing that the strongest correlations between networks is between CC and IA for the three fields. Lower and similar correlations were obtained for CC and IE, and for IE and IA. When communities are considered, the strongest association between communities is between CC and IA networks; the minimum level of association is between IE and CC. 

To the best of our knowledge, the only similar analysis was performed by Ni, Sugimoto and Cronin \citeyear{RN35} in their investigation of scholarly communication. They focused on information and library sciences, by considering networks of journals generated by common authors, co-citation, common topics and common editors. They descriptively compared clusters of journals between networks and calculated a correlation between pairs of matrices by using the quadratic assignment procedure. Their results appears to be coherent with the ones presented here since they estimated statistically significant correlations for networks of journals based on authors, co-citation and editors.

Overall, the results of our analysis show that the intellectual proximity is also a proximity among authors and, more surprisingly, among editors of the journals. This leads to the question of whether the structures obtained could ever be independent if the same set of people were predominantly involved in the editorial boards, the publishing of papers, and the citing of papers. In that case the structures are just a consequence of the existence of a publishing and gatekeeping élite in the considered research fields.\footnote{We owe this observation to one of the two reviewers.} This is a topic worth to be investigated by considering the dual-networks that we used for generating the nine one mode networks analyzed in this paper. At the current state of knowledge, it is only possible to affirm that the map of editorial power, the map of intellectual proximity and the map of author communities tell similar stories. The fact that the results are comparable for the three fields studied suggests that the method presented here is more generally applicable to any scientific field and that there should be in general a coherence among journals at the three scales of 1) editorial boards, 2) authors choice of publications and 3) co-citations. 

\bibliography{dissimilarity}

\begin{thebibliography}{}

\bibitem[\protect\citeauthoryear{Baccini}{Baccini}{2009}]{RN3}
Baccini, A. (2009).
\newblock Italian economic journals. a network-based ranking and an exploratory
  analysis of their influence on setting international professional standards.
\newblock {\em Rivista Italiana degli Economisti\/}~{\em 14\/}(3), 491--511.

\bibitem[\protect\citeauthoryear{Baccini and Barabesi}{Baccini and
  Barabesi}{2010}]{RN21}
Baccini, A. and L.~Barabesi (2010).
\newblock Interlocking editorship. a network analysis of the links between
  economic journals.
\newblock {\em Scientometrics\/}~{\em 82\/}(2), 365--389.

\bibitem[\protect\citeauthoryear{Baccini and Barabesi}{Baccini and
  Barabesi}{2011}]{RN4}
Baccini, A. and L.~Barabesi (2011).
\newblock Seats at the table: The network of the editorial boards in
  information and library science.
\newblock {\em Journal of Informetrics\/}~{\em 5\/}(3), 382--391.

\bibitem[\protect\citeauthoryear{Baccini and Barabesi}{Baccini and
  Barabesi}{2014}]{RN5}
Baccini, A. and L.~Barabesi (2014).
\newblock Gatekeepers of economics: the network of editorial boards in economic
  journals.
\newblock In A.~Lanteri and J.~Vromen (Eds.), {\em The Economics of
  Economists}, pp.\  104--150. Cambridge: Cambridge University Press.

\bibitem[\protect\citeauthoryear{Baccini, Barabesi, and Marcheselli}{Baccini
  et~al.}{2009}]{RN2}
Baccini, A., L.~Barabesi, and M.~Marcheselli (2009).
\newblock How are statistical journal linked? a network analysis.
\newblock {\em Chance\/}~{\em 22\/}(3), 34--43.

\bibitem[\protect\citeauthoryear{Blondel, Guillaume, Lambiotte, and
  Lefebvre}{Blondel et~al.}{2008}]{RN37}
Blondel, V.~D., J.-L. Guillaume, R.~Lambiotte, and E.~Lefebvre (2008, oct).
\newblock Fast unfolding of communities in large networks.
\newblock {\em Journal of Statistical Mechanics: Theory and Experiment\/}~{\em
  2008\/}(10), P10008.

\bibitem[\protect\citeauthoryear{Bretz, Hothorn, and Westfall}{Bretz
  et~al.}{2011}]{RN19}
Bretz, F., T.~Hothorn, and P.~Westfall (2011).
\newblock {\em Multiple Comparisons Using R}.
\newblock Boca Raton: CRC Press.

\bibitem[\protect\citeauthoryear{Brogaard, Engelberg, and Parsons}{Brogaard
  et~al.}{2014}]{RN10}
Brogaard, J., J.~Engelberg, and C.~A. Parsons (2014).
\newblock Networks and productivity: Causal evidence from editor rotations.
\newblock {\em Journal of Financial Economics\/}~{\em 111\/}(1), 251--270.

\bibitem[\protect\citeauthoryear{Cramér}{Cramér}{1946}]{RN17}
Cramér, H. (1946).
\newblock {\em Mathematical Methods of Statistics}.
\newblock Princeton: Princeton University Press.

\bibitem[\protect\citeauthoryear{Crane}{Crane}{1967}]{RN32}
Crane, D. (1967).
\newblock The gatekeepers of science: Some factors affecting the selection of
  articles of scientific journals.
\newblock {\em American Sociologist\/}~{\em 32\/}(2), 195--201.

\bibitem[\protect\citeauthoryear{Csiszar}{Csiszar}{2018}]{RN24}
Csiszar, A. (2018).
\newblock {\em The Scientific Journal. Authorship and Politicis of Knwoledge in
  the Nineteenth Century}.
\newblock Chicago: Chicago University Press.

\bibitem[\protect\citeauthoryear{de~Nooy, Mrvar, and Batagelj}{de~Nooy
  et~al.}{2018}]{RN15}
de~Nooy, W., A.~Mrvar, and V.~Batagelj (2018).
\newblock {\em Exploratory Social Network Analysis with Pajek. Revised and
  expanded edition for updated software}.
\newblock Cambridge: Cambridge University Press.

\bibitem[\protect\citeauthoryear{Dutilleul, Stockwell, Frigon, and
  Legendre}{Dutilleul et~al.}{2000}]{RN20}
Dutilleul, P., J.~D. Stockwell, D.~Frigon, and P.~Legendre (2000).
\newblock The mantel test versus pearson's correlation analysis: Assessment of
  the differences for biological and environmental studies.
\newblock {\em Journal of Agricultural, Biological, and Environmental
  Statistics\/}~{\em 5\/}(2), 131--150.

\bibitem[\protect\citeauthoryear{Erfanmanesh and Morovati}{Erfanmanesh and
  Morovati}{2017}]{RN7}
Erfanmanesh, M. and M.~Morovati (2017).
\newblock Interlocking editorships in scientific journals.
\newblock {\em Science and Engineering Ethics\/}~{\em 2017}.

\bibitem[\protect\citeauthoryear{Heckman and Moktan}{Heckman and
  Moktan}{2018}]{RN6}
Heckman, J.~J. and S.~Moktan (2018).
\newblock Publishing and promotion in economics: The tyranny of the top five.
\newblock {\em National Bureau of Economic Research Working Paper
  Series\/}~{\em No. 25093}.

\bibitem[\protect\citeauthoryear{Hoenig}{Hoenig}{2015}]{RN25}
Hoenig, B. (2015).
\newblock Gatekeepers in social science.
\newblock In J.~D. Wright (Ed.), {\em International Encyclopedia of the Social
  and Behavioral Sciences (Second Edition)}, pp.\  618--622. Oxford: Elsevier.

\bibitem[\protect\citeauthoryear{Hubert and Arabie}{Hubert and
  Arabie}{1985}]{RN18}
Hubert, L. and P.~Arabie (1985).
\newblock Comparing partitions.
\newblock {\em Journal of Classification\/}~{\em 2\/}(1), 193--218.

\bibitem[\protect\citeauthoryear{Klein and DiCola}{Klein and
  DiCola}{2004}]{RN11}
Klein, D.~B. and T.~DiCola (2004).
\newblock Institutional ties of journal of development economics authors and
  editors.
\newblock {\em Econ Journal Watch\/}~{\em 1\/}(2), 319--330.

\bibitem[\protect\citeauthoryear{Kulczycki, Engels, Pölönen, Bruun,
  Dušková, Guns, Nowotniak, Petr, Sivertsen, Istenič~Starčič, and
  Zuccala}{Kulczycki et~al.}{2018}]{RN33}
Kulczycki, E., T.~C.~E. Engels, J.~Pölönen, K.~Bruun, M.~Dušková, R.~Guns,
  R.~Nowotniak, M.~Petr, G.~Sivertsen, A.~Istenič~Starčič, and A.~Zuccala
  (2018).
\newblock Publication patterns in the social sciences and humanities: evidence
  from eight european countries.
\newblock {\em Scientometrics\/}~{\em 116\/}(1), 463--486.

\bibitem[\protect\citeauthoryear{Laband and Piette}{Laband and
  Piette}{1994}]{RN8}
Laband, D.~N. and M.~J. Piette (1994).
\newblock Favoritism versus search for good papers: Empirical evidence
  regarding the behavior of journal editors.
\newblock {\em The Journal of Political Economy\/}~{\em 102\/}(1), 194--203.

\bibitem[\protect\citeauthoryear{Legendre and Legendre}{Legendre and
  Legendre}{1998}]{RN16}
Legendre, P. and L.~Legendre (1998).
\newblock {\em Numerical Ecology}.
\newblock Amsterdam: Elsevier Science.

\bibitem[\protect\citeauthoryear{Lehmann and Romano}{Lehmann and
  Romano}{2005}]{RN34}
Lehmann, E. and J.~Romano (2005).
\newblock {\em Testing Statistical Hypotheses}.
\newblock New York, Springer.

\bibitem[\protect\citeauthoryear{Levandowsky and Winter}{Levandowsky and
  Winter}{1971}]{RN22}
Levandowsky, M. and D.~Winter (1971).
\newblock Distance between sets.
\newblock {\em Nature\/}~{\em 234\/}(5323), 34--35.

\bibitem[\protect\citeauthoryear{Newman}{Newman}{2004}]{RN39}
Newman, M. E.~J. (2004, Jun).
\newblock Fast algorithm for detecting community structure in networks.
\newblock {\em Phys. Rev. E\/}~{\em 69}, 066133.

\bibitem[\protect\citeauthoryear{Newman and Girvan}{Newman and
  Girvan}{2004}]{RN38}
Newman, M. E.~J. and M.~Girvan (2004, Feb).
\newblock Finding and evaluating community structure in networks.
\newblock {\em Phys. Rev. E\/}~{\em 69}, 026113.

\bibitem[\protect\citeauthoryear{Ni, Sugimoto, and Cronin}{Ni
  et~al.}{2013}]{RN35}
Ni, C., C.~R. Sugimoto, and B.~Cronin (2013).
\newblock Visualizing and comparing four facets of scholarly communication:
  producers, artifacts, concepts, and gatekeepers.
\newblock {\em Scientometrics\/}~{\em 94\/}(3), 1161--1173.

\bibitem[\protect\citeauthoryear{Ni, Sugimoto, and Jiang}{Ni
  et~al.}{2013}]{RN36}
Ni, C., C.~R. Sugimoto, and J.~Jiang (2013).
\newblock Venue-author-coupling: A measure for identifying disciplines through
  author communities.
\newblock {\em Journal of the American Society for Information Science and
  Technology\/}~{\em 64\/}(2), 265--279.

\bibitem[\protect\citeauthoryear{Omelka and Hudecova}{Omelka and
  Hudecova}{2013}]{RN28}
Omelka, M. and S.~Hudecova (2013).
\newblock A comparison of the mantel test with a generalised distance
  covariance test.
\newblock {\em Environmetrics\/}~{\em 24\/}(7), 449--460.

\bibitem[\protect\citeauthoryear{Potts, Hartley, Montgomery, Neylon, and
  Rennie}{Potts et~al.}{2017}]{RN9}
Potts, J., J.~Hartley, L.~Montgomery, C.~Neylon, and E.~Rennie (2017).
\newblock A journal is a club: a new economic model for scholarly publishing.
\newblock {\em Prometheus\/}~{\em 35\/}(1), 75--92.

\bibitem[\protect\citeauthoryear{{R Core Team}}{{R Core Team}}{2018}]{RN31}
{R Core Team} (2018).
\newblock {\em R: A Language and Environment for Statistical Computing}.
\newblock Vienna, Austria.

\bibitem[\protect\citeauthoryear{Rizzo and Székely}{Rizzo and
  Székely}{2018}]{RN29}
Rizzo, M. and G.~Székely (2018).
\newblock E-statistics: Multivariate inference via the energy of data.
\newblock {\em R package 'energy'\/}~{\em version 1.7-5}.

\bibitem[\protect\citeauthoryear{Small}{Small}{1973}]{RN26}
Small, H. (1973).
\newblock Co-citation in the scientific literature: A new measure of the
  relationship between two documents.
\newblock {\em Journal of the American Society for Information Science\/}~{\em
  24}, 265--269.

\bibitem[\protect\citeauthoryear{Székely, Rizzo, and Bakirov}{Székely
  et~al.}{2007}]{RN30}
Székely, G.~J., M.~L. Rizzo, and N.~K. Bakirov (2007).
\newblock Measuring and testing dependence by correlation of distances.
\newblock {\em The Annals of Statistics\/}~{\em 35\/}(6), 2769--2794.

\bibitem[\protect\citeauthoryear{Todeschini and Baccini}{Todeschini and
  Baccini}{2016}]{RN1}
Todeschini, R. and A.~Baccini (2016).
\newblock {\em Handbook of Bibliometric Indicators. Quantitative Tools for
  Studying and Evaluating Research}.
\newblock Weinheim (Germany): Wiley-VCH.

\bibitem[\protect\citeauthoryear{Wasserman and Faust}{Wasserman and
  Faust}{1994}]{RN14}
Wasserman, S. and K.~Faust (1994).
\newblock {\em Social Network Analysis: Method and Application}.
\newblock Cambridge: Cambridge University Press.

\end{thebibliography}
\bibliographystyle{chicago}

\newpage

\newpage

\newpage

\end{document}